%Paper: hep-th/9404162
%From: UDAH101@BAY.CC.KCL.AC.UK
%Date: Tue, 26 Apr 94 19:42 BST

%
% NO NUMBER ON FIRST PAGE
%
\def\half {{1 \over 2}}
\def\dzm{{\partial _-}}
\def\dzp{{\partial _+}}
\def\pj{{\partial _j}}

\def\pbj{{\bar\partial _{\bar j}}}
\def\pbk{{\bar\partial _{\bar k}}}

\def\Gbj {{\bar\Gamma^{\bar j}}}
\def\Gbk {{\bar\Gamma^{\bar k}}}
\def\Gbl {{\bar\Gamma^{\bar l}}}
\def\Gj {{\Gamma^j}}
\def\Gk {{\Gamma^k}}
\def\Gl {{\Gamma^l}}

\def\a {{\alpha}}
\def\b {{\beta}}
\def\ad {{\dot\alpha}}
\def\bd {{\dot\beta}}
\def\k {{\kappa}}
\def\kb {{\bar\kappa}}
\def\t {{\theta}}
\def\tb {{\bar\theta}}
\def\ta {{\theta^\alpha}}
\def\tba {{\bar\theta^\ad}}
\def\th {{\hat\theta}}
\def\tbh {{\hat{\bar\theta}}}
\def\N {{\nabla}}
\def\Nb {{\bar\nabla}}
\def\Na {{\nabla_\alpha}}
\def\Nba {{\bar\nabla_\ad}}
\def\D {{D_\k}}
\def\Db {{\bar D_\kb}}

\def\pbad {{\bar p_{\ad}}}
\def\Pa {{\Psi_a}}
\def\Pba {{\bar\Psi_{\bar a}}}

\def\Pbb {{\bar\Psi_{\bar b}}}
\def\Ybj {{\bar Y^{\bar j}}}
\def\ybj {{\bar y^{\bar j}}}
\def\ybk {{\bar y^{\bar k}}}

\def\Ybk {{\bar Y^{\bar k}}}

\tolerance=5000
\footline={\ifnum\pageno>1 \hfil {\rm \folio} \hfil \else \hfil \fi}

\overfullrule=0pt %keeps overfull boxes from showing up in right margin
\baselineskip=18pt
\raggedbottom
\centerline{\bf Covariant Quantization of the Green-Schwarz Superstring}
\centerline{\bf in a Calabi-Yau Background}
\vskip 24pt
\centerline{Nathan Berkovits}
\vskip 24pt
\centerline{Math Dept., King's College,
Strand, London, WC2R 2LS, United Kingdom}
\vskip 12pt
\centerline{e-mail: udah101@kcl.ac.uk}
\vskip 12pt
\centerline {KCL-TH-94-5}
\vskip 12pt
\centerline {April 1994}
\vskip 96pt
\centerline {\bf Abstract}
\vskip 12pt

After adding a scalar chiral boson to the usual superspace variables,
the four-dimensional Green-Schwarz superstring is quantized in a
manifestly SO(3,1) super-Poincar\'e covariant manner. The constraints are
all first-class and form an N=2 superconformal algebra with $c=-3$. Since
the Calabi-Yau degrees of freedom are described by an N=2
superconformal field theory with $c=9$, the combined Green-Schwarz
and Calabi-Yau systems form the $c=6$ matter sector of a critical N=2 string.

Using the standard N=2 super-Virasoro ghosts, a nilpotent BRST charge is
defined and vertex operators for the massless supermultiplets are constructed.
Four-dimensional superstring amplitudes can be calculated with manifest
SO(3,1) super-Poincar\'e invariance by evaluating correlation functions of
these BRST-invariant vertex operators on N=2 super-Riemann surfaces.

\vfil
\eject
\vskip 12pt
\centerline {\bf I. Introduction}
\vskip 12pt

In previous papers by this author,$^{1-4}$
it was shown that the ten-dimensional
Green-Schwarz superstring can be quantized with free fields by
constructing a nilpotent BRST operator out of
an N=2 stress-energy tensor with critical central charge $c=6$.
Scattering amplitudes are calculated by evaluating correlation functions
of BRST-invariant vertex operators on N=2 super-Riemann surfaces, where the
integration over
fermionic super-moduli is performed by inserting N=2 picture-changing
operators. Because this N=2 GS formalism is conformally invariant, it
has the advantage over the light-cone GS formalism$^{5-7}$ that the
picture-changing operators can be inserted anywhere on the surface.
Furthermore, because some of the
spacetime-supersymmetries are manifest, there is no need to perform a GSO
projection or to sum over spin structures as in the RNS formalism.

A disadvantage of the N=2 GS formalism is that the free fields do not
transform linearly under the full set of SO(9,1) super-Poincar\'e
transformations. This prevents the amplitude
calculations from being manifestly
Lorentz-covariant in ten dimensions.
However under an
SO(3,1) subgroup of the super-Poincar\'e transformations,
the free GS fields do transform covariantly.

Under this SO(3,1) subgroup, the ten-dimensional GS fields
split naturally into two groups. The first group describes
a four-dimensional GS superstring which contributes $c=-3$
to the conformal anomaly,
while the second group describes
a flat six-dimensional background with $c=9$.
Since the contribution of the flat six-dimensional background
to the N=2 stress-energy tensor is decoupled from the contribution of the
four-dimensional superstring, the flat background can be replaced by any
Calabi-Yau background characterized by an N=2 superconformal field theory
with $c=9$ (in this paper, the term ``Calabi-Yau''
will mean Ricci-flat Kahler in the large-radius
limit). One therefore has a manifestly SO(3,1)
super-Poincar\'e covariant quantization of the GS
superstring in a Calabi-Yau background.

In Section II of this paper, the worldsheet variables of the
four-dimensional GS superstring are described. In addition to the usual
four-dimensional
superspace variables, there is a scalar chiral boson which is related
both to $R$-transformations in superspace and to projective
transformations in twistor-space. Using these worldsheet variables,
an N=2 stress-energy tensor with central charge
$c=-3$ is constructed
(except for the chiral boson, all of the elements used to construct
this N=2 tensor were described by Siegel in reference 8).
After adding a $c=9$ N=2 superconformal field theory representing the
Calabi-Yau background,
one obtains the $c=6$ matter sector of a critical N=2 string.

In Section III, the N=2 super-Virasoro ghosts are introduced and
a nilpotent BRST charge is defined.
Manifestly SO(3,1) super-Poincar\'e covariant
vertex operators are then
constructed for the massless states of the heterotic
superstring in a Calabi-Yau background.
Unlike in the RNS formalism, all
physical states are represented by vertex operators involving purely
matter fields. In their lowest picture, the vertex operators are constructed
out of prepotentials for the supermultiplets.
Tree-level scattering amplitudes are calculated
in a manifestly SO(3,1) super-Poincar\'e invariant manner by evaluating
correlation functions of these BRST-invariant
vertex operators on N=2 super-Riemann
surfaces.

In Section IV of the paper, the relation between this GS superstring
formalism and the conventional RNS formalism is discussed. It was recently
shown that any critical N=1 string can be embedded in a critical N=2 string,
and the scattering amplitudes coincide using either the N=1 or N=2
prescriptions.$^{9}$ It will be shown in this section
that the GS superstring in a Calabi-Yau background is related
to the RNS string in a Calabi-Yau background by first embedding the N=1
RNS string
in an N=2 string, and then performing a field-redefinition of the worldsheet
variables. This field-redefinition is a conformally invariant version of the
light-cone gauge triality transformation and was described
in reference 10 for the GS superstring in a flat background.

In the final section, possible applications of this paper
are discussed. These include computing multiloop superstring amplitudes,
calculating $\beta$-functions of the GS sigma model, constructing a
GS superstring field theory, and quantizing with manifest SO(9,1)
super-Poincar\'e invariance the
superstring in a flat background.

\vskip 12pt
\centerline {\bf II. The Green-Schwarz Superstring in a Calabi-Yau Background}
\vskip 12pt

\centerline {A. The Four-Dimensional Green-Schwarz Superstring}

The worldsheet variables of the four-dimensional GS superstring
will consist of the spacetime variables, $x^m$ ($m=0$ to 3), the right-moving
fermionic
variables, $\t^\a$ and $\tb^\ad$ ($\a,\ad=1$ to 2), the conjugate right-moving
fermionic variables, $p_\a$ and $\bar p_\ad$, and one right-moving
boson $\rho$. The chiral boson will be defined
so that $\rho$ is identified with $\rho +2\pi$ (in two-dimensional
Minkowski space, $\rho$ is imaginary valued and $i\rho$ takes values
on a circle of radius 1) and will be shown in Section IIIC to be related
to R-transformations of four-dimensional superspace.
For the heterotic GS superstring, one also needs the 32
left-moving chiral
fermions, $\zeta_q$ ($q=1$ to 32), which describe
the SO(32) or $E_8\times E_8$ lattice.
For the four-dimensional Type II GS superstring,
the left-moving fermionic fields,
$\th^\a$,$\tbh^\ad$,
$\hat p_\a$, $\hat{ \bar p}_\ad$, and one left-moving
boson, $\hat\rho$, are needed.

In conformal gauge, the worldsheet action for these fields is:
$$Heterotic: \quad\int dz^+ dz^- [\half\dzp x^m \dzm x_m + p_\a \dzp\t^\a +
\bar p_\ad \dzp\tb^\ad +\half\dzp \rho\dzm\rho +\zeta_q \dzm\zeta_q]
\eqno(1)$$
$$Type~II:\quad
\int dz^+ dz^- [\half\dzp x^m \dzm x_m + p_\a \dzp\t^\a +
\bar p_\ad \dzp\tb^\ad +\half\dzp \rho\dzm\rho +
\hat p_\a \dzm\th^\a +
\hat{\bar p_\ad} \dzm\tbh^\ad +\half\dzp \hat\rho\dzm\hat\rho ].$$
The free-field OPE's for these worldsheet variables are
$$x^m(y) x^n(z)\to -\eta^{mn}\log|y-z|,
\quad \rho(y) \rho(z) \to -\log(y^- -z^-),$$
$$p_\a(y)\theta^\b (z)\to {\delta_\a^\b\over{y^- -z^-}},\quad
\bar p_\ad(y)\bar\theta^\bd (z)\to {\delta_\ad^\bd\over{y^- -z^-}},\quad
\zeta_q(y)\zeta_r(z)\to {\delta_{qr}\over{y^+ -z^+}},$$
$$ \hat p_\a(y)\hat\theta^\b (z)\to {\delta_\a^\b\over{y^+ -z^+}},\quad
\hat{\bar p}_\ad(y)\hat{\bar\theta}^\bd (z)\to
{\delta_\ad^\bd\over{y^+ -z^+}},\quad
\hat\rho(y) \hat\rho(z) \to -\log(y^+ -z^+).$$
Note that the chiral boson $\rho$ can not
be fermionized since
$e^{\rho(y)}~e^{\rho(z)}~\to e^{2\rho(z)}/(y^- -z^-)$ while
$e^{\rho(y)}~e^{-\rho(z)}~\to (y^- -z^-)$. It has the same behavior as the
negative-energy field $\phi$ that appears when bosonizing the RNS ghosts
$\gamma=\eta e^{\phi}$ and $\beta=\partial\xi e^{-\phi}$.$^{11}$

These GS worldsheet variables are constrained by the
$c=-3$ N=2 right-moving
stress-energy tensor:
$$L_{GS}=\half\dzm x^m \dzm x_m +
p_\a\dzm \t^\a + \bar p_\ad \dzm\tb_\ad +\half\dzm\rho\dzm\rho,\quad
G_{GS}=e^{\rho} (d)^2, \quad
\bar G_{GS}=e^{-\rho} (\bar d)^2, \quad
J_{GS}=\dzm\rho,
\eqno(2)$$
where
$$d_\a=p_\a+i\tba\dzm x_{\a\ad}-\half(\tb)^2\dzm\t_\a
+{1\over 4}\t_\a \dzm (\tb)^2,
\quad \bar d_\ad=\bar p_\ad
+i\ta\dzm x_{\a\ad}-\half(\t)^2\dzm\tb_\ad
+{1\over 4}\tb_\ad \dzm (\t)^2,
\eqno(3)$$
and $(d)^2$ means
$\epsilon^{\a\b} d_\a d_\b$. As was shown by Siegel,$^{8}$
$d_\a$ and $\bar d_\ad$ satisfy
the OPE that $d_a(y)$ $d_\b(z)$ is regular,
$d_a (y) \bar d_\ad(z) \to 2i\Pi_{\a\ad}/(y^- -z^-)$ where
$\Pi_{\a\ad}=\sigma^m_{\a\ad}
\dzm x_m -i\t_\a \dzm\tb_\ad-i\tb_\ad\dzm\t_\a,$
and $d_a(y) \Pi^m(z) \to -2i\sigma^m_{\a\ad} \dzm\tb^\ad/(y^- -z^-)$.

It is interesting to note that these four-dimensional GS variables are
closely related to the twistor variables of Penrose$^{12}$ (in fact,
this N=2 description of the GS superstring grew out of the
twistor descriptions of references 13-19). The relation is:
$$\lambda_\a=e^{\rho}d_\a, \quad
\bar\lambda_\ad=e^{-\rho}\bar d_\ad, \quad
\omega^\ad=(x^{\a\ad}+i\ta\tba)e^{\rho}d_\a,\quad
\bar\omega^\a=(x^{\a\ad}-i\ta\tba)e^{-\rho}\bar d_\ad,
\eqno(4)$$
which satisfy the twistor OPE's,
$$\lambda_\a (y)\bar\lambda_\ad (z) \to 2i\Pi_{\a\ad},\quad
\lambda_\a (y)\bar\omega^\b (z) \to {{2i\delta_\a^\b}\over {y^- -z^-}},
\quad
\bar\lambda_\ad (y)\omega^\bd (z) \to {{2i\delta_\ad^\bd}\over {y^- -z^-}}.
\eqno(5)$$
The U(1) current, $J=\dzm\rho$, generates projective transformations in
the twistor-space $CP^3$.

The advantage of working with the variables $d_\a$ and $\Pi^m$ is that they
commute with the spacetime supersymmetry generators,
$$q_\a=\int dz^- [p_\a -i\tba\dzm x_{\a\ad}-{1\over 4}(\tb)^2\dzm\t_\a],
\quad \bar q_\ad=\int dz^- [\bar p_\ad
-i\ta\dzm x_{\a\ad}-{1\over 4}(\t)^2\dzm\tb_\ad].
\eqno(6)$$
Note that the N=2 tensor of equation (2)
is spacetime supersymmetric since $L_{GS}$ can be written in the form
$L_{GS}=\half\Pi^m \Pi_m + d_\a\dzm\t^\a
+ \bar d_\ad \dzm\tb_\ad +\half\dzm\rho\dzm\rho$.

For the heterotic string, the left-moving stress-energy tensor is:
$$\hat L_{GS}=\half\dzp x^m \dzp x_m + \zeta_q \dzp\zeta_q,\eqno(7)$$
while for
the Type II string, the left-moving
N=2 stress-energy tensor is obtained
from equation (2)
by using hatted variables and replacing $\dzm$ with $\dzp$.

To check that this system correctly describes the four-dimensional
GS superstring, one can use the N=2 constraints to gauge-fix to light-cone
gauge. First, use $L_{GS}$ to gauge $\dzm (x^0 +x^3)$ to 1, and use $J_{GS}$
to gauge $\rho$=$i\sigma$ where $\theta^1=e^{i\sigma}$ and
$p_1=e^{-i\sigma}$
(note that
$\rho-i\sigma$ has no singularities with itself).
Since $G_{GS}$ in this gauge contains the term $e^{\rho}p_1 p_2=p_2$,
$G_{GS}$ can be used to gauge $\theta^2=0$. Similarly, $\bar G_{GS}$
in this gauge contains the term
$e^{-\rho}\t^1\bar p_2 \dzm(x^0 +x^3)=\bar p_2$, so $\bar G_{GS}$
can be used
to gauge $\tb^2=0$. Constraining the N=2 tensor to vanish fixes all
variables except for $x^1$, $x^2$, $\tb^1$ and $\bar p_1$, which are just
the light-cone GS variables in four dimensions.

\centerline {B. The Calabi-Yau Background}

The four-dimensional heterotic GS variables described in Section IIA
are related in the following way to the ten-dimensional GS variables
introduced in references 2-4 (for the rest of this paper, only the heterotic
GS superstring will be discussed):
$$x^m=x^m,~~\t^1=\bar\Gamma^4 e^{h^-},~~\t^2=\t^-,~~\tb^1=\Gamma^4 e^{h^+},
{}~~\tb^2=\t^+,
\eqno(8)$$
$$p_1=\Gamma^4 e^{-h^-},~~p_2=\varepsilon^+,~~\bar p_1=\bar\Gamma^4 e^{-h^+},
{}~~\bar p_2=\varepsilon^-,~~\dzm\rho=\dzm(h^- -h^+)+\bar\Gamma^4\Gamma^4.$$
The remaining ten-dimensional GS variables,
$x^\mu$ ($\mu=4$ to 9), $\Gamma^l$ and $\bar\Gamma^{\bar l}$
($l,\bar l=1$ to 3), describe a flat six-dimensional background.
These background variables can be combined in the following way
into N=(2,0) chiral and anti-chiral superfields $Y^j$ and $\Ybj$ for
$j,\bar j=1$ to 3:
$$Y^j(z+\k\kb,\k)=y^j +\kappa \Gamma^j +\k\kb\dzm y^j,\quad
\Ybj(z-\k\kb,\kb)=\ybj +\kb\bar\Gamma^{\bar j}-\k\kb\dzm\ybj,
\eqno(9)$$
where $y^j={1\over{\sqrt{2}}}(x^{j+3}+i x^{j+6})$ and
$\ybj={1\over{\sqrt{2}}}(x^{j+3}-i x^{j+6})$.
The worldsheet action for these variables is
$\int dz^+ dz^- d\k d\kb [ Y^j \dzp \Ybj]$,
while the $c=9$ N=2 right-moving
stress-energy tensor is $T=\D Y^j \Db \Ybj$ where
$\D={\partial\over{\partial\k}} +\kb\dzm$ and
$\Db={\partial\over{\partial\kb}}
 +\k\dzm$. The $c=6$ N=0 left-moving stress-energy
tensor is $\hat L=\dzp Y^j \dzp \Ybj$ at $\k=\kb=0$.

Since the N=2 stress-energy tensor for the flat six-dimensional background
is decoupled from the stress-energy tensor of the four-dimensional
GS superstring, the flat background can be replaced by any $c=9$
N=2 superconformal field theory. As was first shown by Gepner,$^{20}$
$c=9$ N=2 superconformal field theories describe curved
backgrounds which correspond in the large-radius limit to Calabi-Yau
manifolds.

A Calabi-Yau manifold is characterized by a Ricci-flat Kahler metric,
$g_{j\bar k}$, and gauge fields, $A_j$ and $\bar A_{\bar j}$,
whose field strengths satisfy
$F_{jk}$=$F_{\bar j\bar k}$=$g^{j\bar k} F_{j \bar k}=0$.
To reduce the amount of notation, only
the most familiar Calabi-Yau manifold
will be discussed in this paper, namely the manifold described by the
zeros of the polynomial $Z_1^5+Z_2^5+Z_3^5+Z_4^5+Z_5^5$ in $CP^4$ where
the gauge-field is set equal to the spin connection.$^{21}$

The sigma model action for this Calabi-Yau manifold is given by:$^{22}$
$$\int dz^+ dz^- d\k d\kb~ [\pj K~\dzp Y^j -\pbj K~\dzp \Ybj +
\Pa (e^V)^{a \bar b} \Pbb]
\eqno(10)$$
where $K$ is the Kahler potential satisfying
$g_{j\bar k}=\partial_j \bar\partial_{\bar k} K$
and $V^{a\bar b}$
is the gauge prepotential satisfying $\bar A^{a\bar b}_{\bar j}=
(e^{-V})^{a \bar c}(\pbj e^V)^{c\bar b}$
($A^{a\bar b}_j=0$ has been gauge-fixed to zero).
$\Psi_a$ and $\Pba$ for $a,\bar a=1$ to 3
are chiral and
anti-chiral N=(2,0) superfields whose component expansions are
$$\Psi_a (z+\k\kb, \k)=\psi_a +\kappa f_a +\k\kb\dzm\psi_a, \quad
\Pba (z-\k\kb, \kb)=\bar\psi_{\bar a}  +\kb \bar f_{\bar a}-
\k\kb \dzm\bar\psi_{\bar a},
\eqno(11)$$
where $f_a$ and $\bar f_{\bar a}$ are auxiliary bosons,
$\psi_a=\zeta_a +i\zeta_{a+3}$
and $\bar\psi_{\bar a}=\zeta_a -i\zeta_{a+3}$ for $a=1$ to 3,
and $\zeta_q$ for $q=1$ to 6 are chiral fermions
which are ``borrowed'' from the 32
fermions that describe the $E_8\times E_8$ lattice.

In the large-radius limit, the $c=9$ N=2 right-moving
stress-energy tensor is given by
$$T_{CY}= g_{j \bar k}~\D Y^j \Db \Ybk,
\eqno(12)$$
while the $c=9$ N=0 left-moving stress-energy tensor is given by
$$\hat L_{CY}= g_{j \bar k} ~\dzp Y^j \dzp \Ybk
-\dzp Y^j~ \pj(\Psi e^V)^{\bar a}\Pba
+\dzp \Ybj~\Psi_a \pbj (e^V\bar\Psi)^a \quad at~\k=\kb=0.
\eqno(13)$$
Note that the left-moving tensor from the
four-dimensional GS string now contributes $c=17$ to the central
charge since six
chiral fermions have been removed from the 32 $\zeta_q$'s.

The action and stress-tensors for the Calabi-Yau sector
are invariant under the gauge transformations
$$K\to K+\Lambda (Y)+
\bar\Lambda (\bar Y),\quad
e^V\to  e^{\lambda(Y)} e^V e^{\bar\lambda(\bar Y)},~
\Pa\to (\Psi e^{-\lambda})_a,~
\Pba\to (e^{-\bar\lambda}\bar\Psi)_{\bar a},
\eqno(14)$$
where $\pbj\Lambda=\pbj\lambda^{a\bar b}=\pj\bar\Lambda=
\pj\bar\lambda^{a\bar b}=0$.

\vskip 12pt
\centerline {\bf III. Covariant Quantization}
\vskip 12pt

\centerline {A. N=2 Ghosts and the BRST Charge}

By adding together the right-moving
stress-energy tensors of the four-dimensional
GS superstring and of the Calabi-Yau background, one obtains a
$c=6$ N=2 stress-energy tensor which can be coupled to N=2
worldsheet supergravity. The right-moving
N=2 super-Virasoro ghosts that gauge-fix
this coupling consist of the N=(2,0) superfields
$$C=c+\k \gamma+\kb\bar \gamma+\k\kb u,\quad
B=v+\k \bar\beta-\kb \beta+\k\kb b,
\eqno(15)$$
where $(b,c)$ are the usual spin $(2,-1)$ fermionic Virasoro ghosts,
$(\beta,\gamma)$ and $(\bar\beta,\bar\gamma)$ are the two sets of
spin $(3/2,-1/2)$ bosonic superconformal ghosts, and $(v,u)$ are the
spin $(1,0)$ fermionic U(1) ghosts. In terms of these ghost superfields,
the worldsheet action is $\int dz^+ dz^- d\k d\kb [C \dzp B]$ and
the $c=-6$ N=2 stress-energy tensor is
$T=\dzm(CB)-\D C \Db B-\Db C \D B$. It will be convenient to bosonize
the
$(\beta,\gamma)$ and $(\bar\beta,\bar\gamma)$ ghosts in the usual way as
$$\beta=\dzm \xi e^{-\phi},\quad\gamma=\eta e^{\phi},\quad
\bar\beta=\dzm \bar\xi e^{-\bar\phi},\quad
\bar\gamma=\bar\eta e^{\bar\phi}.
\eqno(16)$$

Since the central charge contribution of the ghost fields cancels
the contribution of the matter fields, it is easy to construct a nilpotent
BRST charge as:$^{23}$
$$Q=\int dz^- d\k d\kb [C(T_{GS}+T_{CY})+B(\D C \Db C- C\dzm C)]
\eqno(17)$$
where $T_{GS}$ is the $c=-3$ N=2 stress-energy
tensor of the four-dimensional GS
superstring and $T_{CY}$ is the $c=9$ N=2 stress-energy
tensor of the Calabi-Yau
background.

The left-moving BRST charge, $\hat Q$,
is constructed in the usual way out of
the left-moving Virasoro ghosts $(\hat b,\hat c)$ as
$\hat Q=\int dz^+ [\hat c (\hat L_{GS} +\hat L_{CY} -\dzp \hat c \hat b)]$
where $L_{GS}$ is the
left-moving $c=17$
stress-energy tensor of the four-dimensional heterotic GS string and
$\hat L_{CY}$ is the left-moving $c=9$ stress-energy tensor of the
Calabi-Yau background.

For any critical N=2 string,
it is useful to define
the N=2 picture-changing operators,$^{2}$
$$Z=\{Q,\xi\}= e^{\phi}
[G_{GS}+G_{CY}+ (b-\half\dzm v)\bar\gamma-v\dzm\bar\gamma]+c\dzm\xi,
\eqno(18)$$
$$\bar Z=\{Q,\bar\xi\}= e^{\bar \phi}
[\bar G_{GS}+\bar G_{CY}+ (b+\half\dzm v)\gamma+v\dzm\gamma]+c\dzm\bar\xi,$$
and the instanton-number-changing operators,
$$I= e^{\int^z J_{total}}
= e^{\rho+\phi-\bar\phi+ \int^z J_{CY}},\quad
I^{-1}= e^{-\int^z J_{total}}
=  e^{-\rho-\phi+\bar\phi
-\int^z J_{CY}}.
\eqno(19)$$
Note that these operators have the property that they are BRST-invariant,
but their derivatives are BRST-trivial. Unlike in the N=1 string,$^{11}$
there is
no BRST-invariant
inverse picture-changing operator, $Y$, satisfying $Y(y)$ $Z(z)$=1.

\centerline {B. Vertex Operators}

All physical states of the heterotic GS superstring can be represented
by vertex operators $W$ of the form
$W=c \hat c e^{-\phi-\bar\phi} V$
where $V$ is an N=2 primary field which is constructed entirely out
of matter fields and is dimension (0,1). In other words, $L$, $G$, and
$\bar G$ have only $(y^- -z^-)^{-1}$ singularities with $V$, while $J$ with
$V$ has no singularities. To obtain vertex operators in other pictures,
one can attach arbitrary combinations of $Z$, $\bar Z$, $I$, and $I^{-1}$
onto $W$.

For example, if $V$ depends only on the four-dimensional GS fields and is
independent of the Calabi-Yau manifold,
$$Z \bar Z W= c\hat c~ (\bar d^\ad ~\N^2\Nba V +\dzm\tba~ \Nba V+
\Pi^{\a\ad}~\Na\Nba V)~+ \hat c \gamma~ e^{-\rho}\bar d^\ad ~\Nba V
\eqno(20)$$
where $\Na V= [\int dz^- d_\a~,V]$
and
$\Nba V= [\int dz^- \bar d_\ad~,V]$.
Since $c\hat c$ can be replaced with $\int dz^+ dz^-$, the vertex operator
can be written in integrated form as:$^{24}$
$$\int dz^+ dz^- ~
[\bar d^\ad ~\N^2\Nba V +\dzm\tba~ \Nba V+ \Pi^{\a\ad}~\Na\Nba V].
\eqno(21)$$
Note that this vertex operator is invariant under the gauge transformation
$\delta V=\Lambda +\bar\Lambda$ where $\Nba\Lambda=\Na\bar\Lambda=0$.

The physical
massless states of the heterotic superstring in a Calabi-Yau
background consist of the four-dimensional graviton, gravitino,
axion, dilaton, dilatino, gluon, and gluino, as well as
the modulons and modulinos coming from the marginal deformations
of the Calabi-Yau manifold.
These massless states combine into supermultiplets
which can be represented by BRST-invariant vertex operators in the
following way:

The fields of the supergravity and dilaton multiplets
can be combined into a superfield, $E_m (x,\t,\tb)$, with the gauge
invariances
$\delta E_m=
\Lambda_m +\bar\Lambda_m +\partial_m F$ where
$\Nba\Lambda_m=\Na\bar\Lambda_m=0$. The vertex operator for this superfield
is
$$W=c \hat c e^{-\phi-\bar\phi} ~E_m (x,\t,\tb) ~\dzp x^m ,\eqno(22)$$
which is
BRST-invariant if the following
equations of motion and gauge-fixing conditions are imposed:
$$\N^2 E_m=\Nb^2 E_m=
\eta^{mn} \partial_m E_n=\eta^{mn} \partial_m\partial_n E_p =0.
\eqno(23)$$
The component fields described by this vertex
operator are contained in the $(\t,\tb)$ components of $E_m$.
At $\ta=\tba=0$, the
traceless graviton is $h_{mn}=\sigma_m^{\a\ad}\Na\Nba E_n +
\sigma_n^{\a\ad}\Na\Nba E_m-\half\eta_{mn}\eta^{rs}
\sigma_r^{\a\ad}\Na\Nba E_s $, the dilaton is
$D=\eta^{mn}\sigma_m^{\a\ad}\Na\Nba E_n $, the axion is
$b_{mn}=\sigma_m^{\a\ad}\Na\Nba E_n -
\sigma_n^{\a\ad}\Na\Nba E_m$, the gravitinos are
$\lambda_{m\a}=\Nb^2(\Na E_m -{1\over 8}\sigma_{m\a\ad}\sigma^{n\b\ad}
\nabla_\b E_n$),
$\bar\lambda_{m\ad}=
\N^2(\Nba E_m -{1\over 8}\sigma_{m\a\ad}\sigma^{n\a\bd} \nabla_\bd E_n)$,
and the dilatinos are
$\delta^\a= \N^2(\sigma^{m\a\ad}\Nba E_m)$,
$\bar\delta^\ad= \Nb^2(\sigma^{m\a\ad}\Na E_m)$.
It is easy to check that in Wess-Zumino gauge, the restrictions on $E_m$
from equation (23)
impose the usual polarization and mass-shell conditions on these component
fields (e.g. $\partial_m h^{mn}=\partial_n h^{mn}=\eta^{mn}\partial_m
\partial_n h^{rs}=0$).

The super-Yang-Mills fields
can be combined into a superfield, $V^I(x,\t,\tb)$,
with the gauge invariance
$\delta V^I=
\Lambda^I +\bar\Lambda^I$ where $\Nba\Lambda^I=\Na\bar\Lambda^I=0$ and
$I$ takes values in the adjoint
representation of the unbroken
$E_6\times E_8$. The vertex operator for this superfield
is
$$W=c \hat c e^{-\phi-\bar\phi} ~V^I (x,\t,\tb) ~j_I\eqno(24)$$
where $j_I$ is the current
for $E_6\times E_8$
that is constructed out of the 26 chiral fermions
($j_I$ is quadratic in the chiral fermions for an
$SO(10)\times SO(16)$ subgroup of $E_6\times E_8$,
but for the other elements of $E_6\times E_8$, $j_I$ is
realized non-linearly). This vertex operator is BRST-invariant
after imposing the equations of motion and gauge-fixing conditions:
$$\N^2 V^I=\Nb^2 V^I= \eta^{mn}\partial_m\partial_n V^I =0.
\eqno(25)$$
At $\ta=\tba=0$, the
gluon is $A_m^I=\sigma_m^{\a\ad}\Na \Nba V^I$, and the gluinos are
$\chi_\a^I=\Nb^2\Na V^I$, $\bar \chi_\ad^I=\N^2\Nba V^I$.

Each marginal deformation of the Calabi-Yau manifold gives rise to a
massless complex scalar and Weyl spinor in four dimensions. These combine
into chiral and anti-chiral superfields,
$\Omega (x-i\t\sigma\tb,\t)$ and
$\bar\Omega (x+i\t\sigma\tb,\tb)$,
satisfying $\Nba\Omega=\Na\bar\Omega=0$.
On-shell, these superfields
satisfy the equations of motion
$\N^2 \Omega=\Nb^2\bar\Omega=0$.
In order to write the exact vertex operators for $\Omega$ and
$\bar\Omega$,
one needs to know the primary fields of the $c=9$ N=2 superconformal
field theory. However in the large-radius limit, these vertex operators
can be approximated by expressions involving the Calabi-Yau fields.

For the massless states that come from deforming the Kahler structure
of the Calabi-Yau manifold, the vertex operators are
$$W
=c \hat c e^{-\phi-\bar\phi}~ \Omega~ \pj K ~\dzp y^j,\quad
\bar W=
c \hat c e^{-\phi-\bar\phi}~\bar\Omega ~\pbj K~\dzp \ybj,
\eqno(26)$$
where $K$ is the Kahler potential satisfying
$\pj \pbk K=g_{j\bar k}$. Note that
these vertex operators change by a BRST-trivial
quantity under the gauge transformation $\delta K=\Lambda
+\bar\Lambda$ where $\pbj\Lambda=\pj\bar\Lambda=0$, and that in the picture
$\bar Z W$ and
$Z \bar W$, this gauge invariance is manifest since
$$\bar Z W
=c \hat c e^{-\phi} ~\Omega ~g_{j\bar k}~\Gbk \dzp y^j ,\quad
Z \bar W=
c \hat c e^{-\bar\phi}~\bar\Omega ~g_{j \bar k}~\Gj\dzp \ybk.
\eqno(27)$$
By introducing potentials for the one-forms on the
Calabi-Yau manifold, it is also possible to write the
vertex operators for the other massless states in
the form $W=c \hat c e^{-\phi-\bar\phi} \Omega V$ and
$\bar W=c \hat c e^{-\phi-\bar\phi} \bar\Omega \bar V$. However, it is
more
convenient to write them in the picture $\bar Z W$ and $Z\bar W$ so that
the one-forms appear directly.

For the massless states
that come from deforming the complex structure of the manifold,
the vertex operators are
$$\bar Z W= c \hat c e^{-\phi}~\Omega ~g_{j\bar k}
\bar K_{\bar l}^j ~\Gbl\dzp \ybk,\quad
Z\bar W= c \hat c e^{-\bar\phi}~\bar\Omega~
g_{k\bar j} K_l^{\bar j}~\Gl \dzp y^k,
\eqno(28)$$
where $\bar K_{\bar l}^{j}$
and $K_l^{\bar j}$
are the 101 complex elements in the Dolbeault cohomology
$H^1(T)$ and $\bar H^1(\bar T)$ ($T$ and $\bar T$ are the holomorphic
and anti-holomorphic tangent bundles of the manifold).

Marginal deformations of the Calabi-Yau gauge
field give rise to massless states which are
singlets, $27$'s and $\bar {27}$'s of the $E_6\times E_8$.
The vertex operators for the singlets are
$$\bar Z W=c \hat c e^{-\phi}~ \Omega ~\bar M_{\bar j}^{a\bar b}~\Gbj
\psi_a\bar\psi_{\bar b} ,\quad
Z \bar W=
c \hat c e^{-\bar\phi}~\bar\Omega ~
M_j^{a\bar b}~\Gj\psi_a\bar\psi_{\bar b}
\eqno(29)$$
where $\bar M_{\bar j}^{a\bar b}$
and $M_j^{a\bar b}$ are the
224 complex elements in the Dolbeault cohomology $H^1(End ~T)$ and
$\bar H^1(End~ \bar T)$. The vertex operators for the $27$'s are
$$\bar Z W=c \hat c e^{-\phi}~ \Omega^u~ \bar M_{\bar k}^a ~\Gbk
j_{(u,a)},\quad
Z \bar W=
c \hat c e^{-\bar\phi}~\bar\Omega^u~
M_k^a~\Gk j_{(u,a)}
\eqno(30)$$
where $\bar M_{\bar k}^a$ are the 101 elements in
$H^1(T)$, $M_k^a$ is the unique element in $\bar H^1 (T)$,
$u$ is the label for the $27$ representation, and $j_{(u,a)}$ are the
81 currents constructed out of the $\zeta_q$'s which form the (27,3)
in the decomposition of the adjoint representation of $E_8$. The vertex
operators for the $\bar {27}$'s are the complex conjugates of the vertex
operators for the $27$'s.

\centerline {C. Scattering Amplitudes}

Scattering amplitudes for this N=2 GS formalism are calculated in the same
way as for any critical N=2 string.$^{25}$
Correlations of BRST-invariant vertex
operators are evaluated on an N=2 super-Riemann surface, and the moduli and
super-moduli of the N=2 surface are integrated over. Integration over the
fermionic super-moduli is straightforward and gives rise to insertions of
the picture-changing operators, $Z$ and $\bar Z$, of equation (18).
Integration over the U(1) moduli,
however, is more subtle due to complications in
defining the path integral over the negative-energy $\rho$ field.$^{2}$
To avoid these subtleties, only tree-level scattering amplitudes will be
discussed here, although
another paper will hopefully be written soon discussing the
loop amplitude calculations.

On a sphere with $N$ punctures, there
are no U(1) moduli, however there is still an instanton number, $n_I$,
coming from the integral of the field-strength of the U(1) gauge field.
For the heterotic string, the number of bosonic moduli is $2N-6$, the
number of fermionic moduli coming from $G$ is $N-2+n_I$, and the number
of fermionic moduli coming from $\bar G$ is $N-2-n_I$ (for $|n_I|>N-2$,
the amplitude vanishes).$^{2}$
Evaluating a correlation function on a surface with instanton number
$n_I$ is equivalent to evaluating it on a surface with instanton number zero
and inserting $n_I$ instanton-number-changing operators $I$ from eqn. (19).
After choosing the bosonic moduli to be located at $N-3$ of the punctures
and after integrating over the fermionic moduli, the tree-level
scattering amplitude for N external states $V_1,...,V_N$ is given by:
$$A_{V_1,...,V_N}=
\sum_{n_I=2-N}^{N-2}
 < \hat c c e^{-\phi-\bar\phi} V_1(z_1)~
\hat c c e^{-\phi-\bar\phi} V_2(z_2)~
\hat c c e^{-\phi-\bar\phi} V_3(z_3)
\eqno(31)$$
$$
\int d^2 z_4 e^{-\phi-\bar\phi} V_4(z_4)~ ...~
\int d^2 z_N e^{-\phi-\bar\phi} V_N(z_N)~~
I^{n_I} ~Z^{N-2+n_I} ~\bar Z^{N-2-n_I}>$$
where the locations of the $I$'s, $Z$'s and $\bar Z$'s are arbitrary.
Because of the background charges of the various fields, the correlation
function
$$<(\t)^2 (\tb)^2~ c \dzm c \partial_-^2 c ~\hat c \dzp \hat c
\partial_-^2 \hat c
{}~e^{-2\phi-2\bar\phi}>=1.
\eqno(32)$$
The presence of the $(\t)^2 (\tb)^2$ zero modes
guarantees that the amplitude is SO(3,1) super-Poincar\'e invariant.

It is interesting to note that for tree amplitudes involving external
states which are independent of the Calabi-Yau fields, only surfaces with
$n_I=0$ contribute. This can be seen by observing that $I$ explicitly
depends on the Calabi-Yau fields through the term $e^{-\int^z
J_{CY}}$. Although
the $Z$'s and $\bar Z$'s also depend on the Calabi-Yau fields, there are
not enough $Z$'s and $\bar Z$'s to cancel the dependence of the $I$'s
(this result is not valid for loop amplitudes since there are more
$Z$'s and $\bar Z$'s present).
As will now be shown, this property of tree amplitudes is caused by
$R$-invariance of the classical four-dimensional
action.

In four-dimensional superspace, the $R$-transformation scales $\ta\to
e^{ir} \ta $, $\tba\to e^{-ir}\tba$, and $\Omega\to e^{iwr}\Omega$
where $w$ is the $R$-weight of the four-dimensional superfield $\Omega$.
In the GS superstring, this transformation is generated by the BRST-invariant
operator, $R=\int dz^- (2\dzm \rho-\ta p_a +\tba \pbad)$, where
$[\dzm\rho,\Omega]=-\half w\Omega$ (the $R$-weights of the
super-Yang-Mills prepotential, $V$, and of the supergravity/dilaton
potential, $E_m$, are zero since the vertex operators have no $e^{\rho}$
factors).
Note that $[R,Z]=[R,\bar Z]=0$, but $[R,I]=-2I$.

If $R$-invariance is a symmetry of the classical four-dimensional action,
the tree amplitude
$A_{[R,V_1 ... V_N]}=A_{[R,V_1] V_2 ... V_N}+ ... +
A_{V_1 ... V_{N-1} [R,V_N]}=0.$ But by pulling the contour of
$\int dz^- (2\dzm \rho-\ta p_a +\tba \pbad)$ around the various operators in
expression (31) for the scattering amplitude, this implies that
$\sum_{n_I=2-N}^{N-2} n_I A_{n_I}=0$ where
$$A_{n_I}=
<\hat c c e^{-\phi-\bar\phi} V_1(z_1)~
\hat c c e^{-\phi-\bar\phi} V_2(z_2)~
\hat c c e^{-\phi-\bar\phi} V_3(z_3)
\eqno(33)$$
$$
\int d^2 z_4 e^{-\phi-\bar\phi} V_4(z_4)~ ...~
\int d^2 z_N e^{-\phi-\bar\phi} V_N(z_N)~~
I^{n_I}~ Z^{N-2+n_I}~ \bar Z^{N-2-n_I}>.$$ Similarly,
$A_{[R,[R,V_1 ... V_N]]}=0$ implies that
$\sum_{n_I=2-N}^{N-2} n_I^2  A_{n_I}=0$.
By repeatedly commuting with
$R$, one therefore proves that $A_{n_I}=0$ for $n_I\neq 0$.

\vskip 12pt
\centerline {\bf IV. Relation with the RNS Formalism}
\vskip 12pt

It was recently shown that any critical N=1 string can be embedded in a
critical N=2 string and the scattering amplitudes coincide using
the N=1 and
N=2 prescriptions.$^{9}$ It was also shown that the BRST operators
in the N=1 and N=2 formalisms are related by a
similarity transformation.$^{26}$
If one chooses the critical N=1 string to be an RNS string in a Calabi-Yau
background, the right-moving $c=6$ N=2 stress-energy tensor is:
$$L=\half\dzm x^m\dzm x_m +\psi^m \dzm \psi_m + L_{CY}^{RNS} -{3\over 2}
\beta\dzm\gamma-{\half\gamma\dzm\beta}-{3\over 2}b\dzm c +\half c\dzm b
+\half \dzm(\xi\eta),$$
$$G=b, \quad\quad\quad
\bar G=\gamma (\psi^m \dzm x_m + G_{CY}^{RNS} +\bar G_{CY}^{RNS}) +
\eqno(34)$$
$$c(\half\dzm x^m\dzm x_m +\psi^m \dzm \psi_m + L_{CY}^{RNS} -{3\over 2}
\beta\dzm\gamma-{\half\gamma\dzm\beta}-b\dzm c )-\gamma^2 b +\partial_-^2 c
+\dzm (c\xi\eta),$$
$$J=cb+\eta\xi,$$
where $x^m$ and $\psi^m$ for $m=0$ to 3 are the four-dimensional RNS
matter fields, $[b,c]$ and
$[\beta=\dzm \xi e^{-\phi},\gamma=\eta e^{\phi}]$
are the twisted RNS ghost fields, and $[L_{CY}^{RNS},
G_{CY}^{RNS},\bar G_{CY}^{RNS}, J_{CY}^{RNS}]$ is the $c=9$ N=2
stress-energy tensor $T_{CY}^{RNS}$
for the RNS Calabi-Yau background (as will be shown
in equation (36), $T_{CY}^{RNS}$ is related by a field
redefinition to the $c=9$ N=2 stress-energy tensor of the GS Calabi-Yau
background, $T_{CY}$ of equation (12)).

In reference 10, it was shown that for a flat background, the N=2
tensor constructed out of the ten-dimensional
RNS matter and ghost fields is mapped
by a field redefinition onto the N=2 tensor for the ten-dimensional
GS fields (this was how the N=2 RNS tensor was originally found).
It is easy to modify this field redefinition
for a curved Calabi-Yau background so that it
maps the N=2 RNS tensor of equation (34) onto the N=2 GS tensor
$T_{GS}+T_{CY}$ of equations (2) and (12).

The first step is to define a ``chiral'' set of GS variables by
performing the unitary transformation,
$$\tilde\Phi=e^{-\int dz^- [i\dzm x_{\a\ad} \ta\tba
+ e^{-\rho}(\t)^2 G_{CY}]}~
\Phi~ e^{\int dz^- [i\dzm x_{\a\ad} \ta\tba
+ e^{-\rho}(\t)^2 G_{CY}]},
\eqno(35)$$
where $\Phi$ includes all GS fields defined in Section II. In terms of
these chiral GS variables, $G_{GS}+G_{CY}$ is
simply $e^{\tilde\rho} (\tilde p)^2$.
The field redefinition from these chiral GS variables to the RNS
variables is:
$$\tilde x^m_{GS}=x^m_{RNS},\quad
\dzm\tilde\rho=-3\dzm\phi+cb+2\xi\eta-J_{CY}^{RNS},
\eqno(36)$$
$$ \tilde\t^1=c \xi e^{\half(-3\phi+
\int^z [-\psi^0\psi^1+\psi^2\psi^3-J_{CY}^{RNS}])},
\quad
\tilde\t^2=c \xi e^{\half(-3\phi+
\int^z[\psi^0\psi^1-\psi^2\psi^3-J_{CY}^{RNS}])},$$
$$\tilde\tb^1=e^{\half(\phi+
\int^z[-\psi^0\psi^1-\psi^2\psi^3+J_{CY}^{RNS}])},
\quad
\tilde\tb^2=e^{\half(\phi+
\int^z[\psi^0\psi^1+\psi^2\psi^3+J_{CY}^{RNS}])}$$
$$\tilde{p_1}=b \eta e^{\half(3\phi+
\int^z[\psi^0\psi^1-\psi^2\psi^3+J_{CY}^{RNS}])},
\quad
\tilde{p_2}=b \eta e^{\half(3\phi+
\int^z[-\psi^0\psi^1+\psi^2\psi^3+J_{CY}^{RNS}])},$$
$$\tilde{\bar p_1}=e^{\half(-\phi+
\int^z[\psi^0\psi^1+\psi^2\psi^3-J_{CY}^{RNS}])},
\quad
\tilde {\bar p_2}=e^{\half(-\phi+
\int^z[-\psi^0\psi^1-\psi^2\psi^3-J_{CY}^{RNS}])},$$
$$\tilde L_{CY}=L_{CY}^{RNS}+{3\over 2}(\dzm\phi+\eta\xi)^2-
(\dzm\phi+\eta\xi)J_{CY}^{RNS},\quad
\tilde G_{CY}
=e^{\phi}\eta G_{CY}^{RNS}$$
$$ \tilde{\bar G}_{CY}
=e^{-\phi}\xi \bar G_{CY}^{RNS},
\quad
\tilde J_{CY}=J_{CY}^{RNS}+3(\dzm\phi+\eta\xi).$$

Note that any GS field can be transformed to an RNS field, but
only GSO-projected RNS fields (i.e., fields which have no square-root
cuts with the spacetime-supersymmetry generator)
can be transformed into single-valued GS fields. In addition to mapping
the N=2 GS tensor $T_{GS}+T_{CY}$ onto the N=2 RNS tensor of equation (34),
this field redefinition maps the integrated GS vertex operators
(in the picture $Z\bar Z W$ where they have no N=2 ghost dependence) onto
integrated RNS vertex operators and maps the GS spacetime-supersymmetry
generators of eqn. (6) onto the RNS spacetime-supersymmetry generators
$$q_1 =
 \int dz^- [~b\eta e^{\half(3\phi+
\int^z[\psi^0\psi^1-\psi^2\psi^3+J_{CY}^{RNS}])}
\eqno(37)$$
$$-e^{\phi\over 2}~
\{~\psi_m\dzm x^m+ G_{CY}^{RNS}+\bar G_{CY}^{RNS} ~,~
e^{\half\int^z[\psi^0\psi^1-\psi^2\psi^3+J_{CY}^{RNS}]}~\}~],$$
$$q_2 =
\int dz^- [~b\eta e^{\half(3\phi+
\int^z[-\psi^0\psi^1+\psi^2\psi^3+J_{CY}^{RNS}])}$$
$$-e^{\phi\over 2}~
\{~\psi_m\dzm x^m+ G_{CY}^{RNS}+\bar G_{CY}^{RNS}~,~
e^{\half\int^z [-\psi^0\psi^1+\psi^2\psi^3+J_{CY}^{RNS}]}~\}~],$$
$$\bar q_1 = \int dz^- e^{\half(-\phi+
\int^z[\psi^0\psi^1+\psi^2\psi^3-J_{CY}^{RNS}])},$$
$$ \bar q_2 = \int dz^- e^{\half(-\phi+
\int^z[-\psi^0\psi^1-\psi^2\psi^3-J_{CY}^{RNS}])}.  $$

Since $R=
\int dz^- (2\dzm \rho-\ta p_\a +\tba \bar p_\ad)$ gets mapped onto the RNS
operator
$\int dz^- (2\xi\eta-2\dzm\phi)$,
the $R$-weight of a GS operator is equal to twice the picture of the
corresponding RNS operator. In other words, the vertex operators for
chiral fermions in the GS formalism get mapped onto Ramond states in the
$+\half$ picture, while the vertex operators for anti-chiral fermions
get mapped onto Ramond states in the $-\half$ picture. In the N=1
RNS formalism,
this identification of chirality and ghost number would be inconsistent
since it would force amplitudes to vanish unless they had a fixed number of
chiral minus anti-chiral external states. In the N=2 formalism, however,
there is no inconsistency because of the sum over instanton number which
picks up contributions from different RNS pictures.

\vskip 12pt
\centerline {\bf V. Possible Applications}
\vskip 12pt

There are various longstanding problems in superstring theory for
which the results of this paper might be useful. One such problem is
to calculate multiloop superstring amplitudes in a manifestly SO(3,1)
super-Poincar\'e invariant manner.
In previous papers by this author,$^{2-4}$
multiloop GS amplitudes were calculated
in a flat background using variables which were not manifestly SO(3,1)
covariant. The advantage of using non-covariant variables is that it is
easy to compare with light-cone gauge calculations to check that the
amplitudes are unitary.$^{4}$ Because the rules for integrating over the U(1)
moduli are not straightforward, this check of unitarity is especially
important.

Of course, the advantage of using covariant variables is that the
amplitude calculations would be manifestly SO(3,1) super-Poincar\'e
invariant. However, proving equivalence with light-cone calculations
may be tricky since, as was described at the end of Section IIA,
light-cone gauge fixing using the covariant variables involves bosonization.

Another problem which can be re-evaluated using the techniques of this
paper is to derive the superstring corrections to the classical equations
of motion for the supergravity and super-Yang-Mills fields.
This is done by coupling the GS superstring to a curved four-dimensional
background and requiring that the resulting non-linear sigma
model is conformally invariant. Previous attempts$^{27,28}$ to derive the
superstring corrections used semi-light-cone gauge fixing$^{29}$ of the
fermionic symmetries, which led to incomplete results in four
dimensions.$^{27}$ Since an
N=(2,0) non-linear sigma model for the superstring in a curved
background has already been constructed,$^{30,31}$
it should be straightforward
to use the N=(2,0) techniques described in this paper to derive the
superstring corrections to the massless equations of motion.

A third possible application of this paper is to develop a manifestly SO(3,1)
super-Poincar\'e invariant superstring field theory. Although the role of
the chiral boson, $\rho$, needs to be better understood, it should
be possible to construct a field theory action of the form
$<\Phi Q\Phi + Z\bar Z\Phi^3 +...>$.$^{32}$
As in the RNS string field theory,
contact terms will be necessary to cancel the divergences of
colliding picture-changing operators, $Z$ and $\bar Z$.$^{33,34}$

An obvious question is if it is possible to quantize the superstring in
a flat background and preserve all SO(9,1) super-Poincar\'e invariance.
A major obstacle to this goal is that the sixteen $\t$ variables of
SO(9,1) superspace are not free fields. The easiest way to see this is that
in the RNS formalism, $\ta =e^{{\phi}\over 2} S^\a$ where $S^\a$ is
the spin field constructed by bosonizing the ten $\psi^\mu$'s.$^{11}$ But
$\ta (y) \theta^\b (z) \to (y^- -z^-)^{-1} e^{\phi}\gamma_\mu^{\a\b}\psi^\mu$
which is not a free-field operator product. Note that the maximum
number of $\t$'s which have no singularities with each other is five,
which transform as $1_1$ and $4_\half$ representations under the
subgroup SU(4)$\times$U(1) of SO(9,1). So unless some new description
of ten-dimensional superspace is constructed in which the $\t$'s are
not fundamental fields, it seems unlikely that the superstring
will be quantizable in a manifestly SO(9,1) super-Poincar\'e invariant
manner.

It is interesting to note that the most
natural description of the GS superstring in ten flat dimensions
contains N=8 worldsheet supersymmetry,$^{35-38}$ rather than N=2. Since the
superstring description of this paper can be obtained by embedding the
N=1 RNS string into an N=2 string, perhaps one should try to embed
the RNS string into an N=8 string. It is possible that a new picture
of ten-dimensional superspace could emerge from the resulting
N=8 description of the superstring.

\vskip 42pt
\centerline {\bf Acknowledgements}
\vskip 12pt
I would like to thank P. Howe, S. Kachru, P. van Nieuwenhuizen,
M. Ro\v cek, E. Sezgin,
W. Siegel, C. Vafa, E. Witten
for useful discussions, and the SERC for its financial support.
I would also like to thank the ITP at Stony Brook where part of this
work was completed.

\vskip 12pt

\centerline{\bf References}
\vskip 12pt

\item{(1)} Berkovits,N., Nucl.Phys.B379 (1992), p.96., hep-th no. 9201004.

\item{(2)} Berkovits,N., Nucl.Phys.B395 (1993), p.77, hep-th no. 9208035.

\item{(3)} Berkovits,N., Phys.Lett.B300 (1993), p.53, hep-th no. 9211025.

\item{(4)} Berkovits,N., Nucl.Phys.B408 (1993), p.43, hep-th no. 9303122.

\item{(5)} Green,M.B. and Schwarz,J.H., Nucl.Phys.B243 (1984), p.475.

\item{(6)} Mandelstam,S., Prog.Theor.Phys.Suppl.86 (1986), p.163.

\item{(7)} Restuccia,A. and Taylor,J.G., Phys.Rep.174 (1989), p.283.

\item{(8)} Siegel,W., Nucl.Phys.B263 (1986), p.93.

\item{(9)} Berkovits,N. and Vafa,C., Mod.Phys.Lett.A9 (1994), p.653,
hep-th no. 9310129.

\item{(10)} Berkovits,N., ``The Ten-Dimensional Green-Schwarz Superstring
is a Twisted Neveu-Schwarz-Ramond String'', KCL-TH-93-12 (Aug. 1993),
to appear in
Nucl.Phys.B, hep-th no. 9308129.

\item{(11)} Friedan,D., Martinec,E., and Shenker,S., Nucl.Phys.B271
(1986), p.93.

\item{(12)} Penrose,R. and MacCallum,M.A.H., Phys.Rep.6C (1972), p.241.

\item{(13)} Ferber,A., Nucl.Phys.B132 (1978), p.55.

\item{(14)} Witten,E., Nucl.Phys.B266 (1986), p.245.

\item{(15)} Bengtsson,I. and Cederwall,M., Nucl.Phys.B302 (1988), p.81.

\item{(16)} Sorokin,D.P., Tkach,V.I., Volkov,D.V., and Zheltukhin,A.A.,
Phys.Lett.B216 (1989), p.302.

\item{(17)} Berkovits,N., Phys.Lett.B232 (1989), p.184.

\item{(18)} Tonin,M., Phys.Lett.B266 (1991), p.312.

\item{(19)} Ivanov,E.A. and Kapustnikov,A.A., Phys.Lett.B267 (1991), p.175.

\item{(20)} Gepner,D., Phys.Lett.B199 (1987), p.380.

\item{(21)} Candelas,P., Horowitz,G., Strominger,A. and Witten,E.,
Nucl.Phys.B258 (1985), p.46.

\item{(22)} Brooks,R., Gates,S.J.Jr. and Muhammed,F., Nucl.Phys.B268
(1986), p.599.

\item{(23)} Giveon,A. and Ro\v cek,M., Nucl.Phys.B400 (1993), p.145.

\item{(24)} Mikovic,A., Preitschopf,C. and van de Ven,A., Nucl.Phys.B321
(1989), p.121.

\item{(25)} Ooguri,H. and Vafa,C., Nucl.Phys.B361 (1991), p.469.

\item{(26)} Ohta,N. and Petersen,J.L., ``N=1 from N=2 Superstrings''.
NBI-HE-93-76 (Jan. 1994), hep-th no. 9312187.

\item{(27)} Gates,S.J.Jr., Majumdar,P., Oerter,R.N. and van de Ven,A.E.,
Phys.Lett.B214 (1988), p.26.

\item{(28)} Grisaru,M., Nishino.H. and Zanon,D., Nucl.Phys.B314 (1989),
p.363.

\item{(29)} Carlip,S., Nucl.Phys.B284 (1987), p.365.

\item{(30)} Delduc,F. and Sokatchev,E., Class.Quant.Grav.9 (1992), p.361.

\item{(31)} Berkovits,N., Phys.Lett.B304 (1993), p.249, hep-th no. 9303025.

\item{(32)} Witten,E., Nucl.Phys.B276 (1986), p.291.

\item{(33)} Greensite,J. and Klinkhamer,F.R., Nucl.Phys.B291 (1987), p.557.

\item{(34)} Wendt,C., Nucl.Phys.B314 (1989), p.209.

\item{(35)} Berkovits,N., Phys.Lett.B241 (1990), p.497.

\item{(36)} Berkovits,N., Nucl.Phys.B358 (1991), p.169.

\item{(37)} Delduc,F., Galperin,A., Howe,P., and Sokatchev,E.,
Phys.Rev.D47 (1993), p.578.

\item{(38)} Brink,L., Cederwall,M. and Preitschopf,C., Phys.Lett.B311 (1993),
p.76.

\end